\documentclass[conference]{IEEEtran}
\usepackage{cite}

\ifCLASSINFOpdf
  \usepackage[pdftex]{graphicx}
\else
\fi
\usepackage{amsmath,amssymb,amsfonts}
\usepackage{algorithmic}
  \usepackage[caption=false,font=footnotesize]{subfig}
\usepackage{url}

\usepackage{xcolor}

\usepackage{eurosym}
\DeclareRobustCommand{\officialeuro}{%
  \ifmmode\expandafter\text\fi
  {\fontencoding{U}\fontfamily{eurosym}\selectfont e}}

\begin{document}
%
\title{Impact of Grid Tariffs Design on the Zero Emission Neighborhoods’ Energy System Investments}

\author{\IEEEauthorblockN{Dimitri Pinel, Sigurd Bjarghov, Magnus Korp{\r a}s}
\IEEEauthorblockA{Dept. of Electric Power Engineering\\
Norwegian University of Science and Technology\\
Trondheim, Norway\\
dimitri.q.a.pinel@ntnu.no, 
sigurd.bjarghov@ntnu.no, 
magnus.korpas@ntnu.no}
}

\maketitle

\begin{abstract}
This paper investigates the relationship between grid tariffs and investment in Zero Emission Neighborhoods (ZEN) energy system, and how the grid exchanges are affected. Different grid tariffs (energy based, time of use (ToU), subscribed capacity and dynamic) are implemented in an optimization model that minimizes the cost of investing and operating a ZEN during its lifetime. The analysis is conducted in two cases: non-constrained exports and exports limited to 100kWh/h. The results suggest that in the case with no limit on export, the grid tariff has little influence, but ToU is economically advantageous for both the ZEN and the DSO. When exports are limited, the subscribed capacity scheme allows to maintain DSO revenue, while the others cut them by half. This tariff also offers the lowest maximum peak and a good duration curve. The dynamic tariff creates new potentially problematic peak imports despite its benefits in other peak hours.
\end{abstract}
\vspace{6pt}

\IEEEpeerreviewmaketitle

\section{Introduction}
The structure of grid tariffs has recently become a more important topic due to the increasing amount of prosumers in the grid and a large implementation of smart meters enabling more complex price structures than is common today. Policy makers, transmission system operators (TSO) and distribution system operators (DSO), need to assess the benefits and drawbacks associated with changing the traditional energy based grid tariffs into more complex formulations such as capacity subscription, time-of-use tariffs, real-time pricing etc. Some of the expected benefits would include reducing grid expansion, peak loads and/or incentivizing end-user flexibility while drawbacks could be a less intuitive pricing structure for consumers or unfairness due to cross-subsidization. 

Grid tariffs have specific requirements to meet. They are supposed to reflect the cost of the maintenance, losses and in some cases the cost of grid expansion necessary for new connections \cite{picciariello2015}. Grid tariffs are made up of one or several of the following components:
\begin{itemize}
    \item a fixed part paid typically each month or year, independently of the utilization of the grid ($\euro$)
    \item an energy part, based on the amount of energy consumed ($\euro/kWh$)
    \item a power part, based on either a subscribed capacity or the size of the connection ($\euro/kW$)
\end{itemize}
Variations around these structures can be made by taking into account additional parameters such as time or several power levels for example.
Several principles are often mentioned when it comes to how the tariff should be. They aim at having a sustainable economically efficient system while protecting consumers \cite{picciariello2015}. In more detail, the system should guarantee universal access to electricity with a transparent, simple, stable and equitable pricing system representing each user's contribution to the cost and allowing the grid company to recover the total cost while maintaining it as low as possible  \cite{picciariello2015}, \cite{herbst2016}.

In Norway, the grid tariff varies depending on the region, with more remote areas paying a higher grid tariff. On average for households the tariffs are: a fixed part of 181$\euro/year$ and a variable part of $0,020\euro/kWh$ \cite{NordREG2015}. References \cite{NordREG2015} and \cite{optimal_2008} also details the law the grid tariffs have to abide by, the situation of the tariffs and the trend in Norway to move towards more power based tariffs in the future.

In parallel, Zero Emission Neighborhoods (ZEN) is a concept being developed in Norway in the ZEN research center and follows the work of the research center for Zero Emission Buildings. ZENs are neighborhoods that reduce their greenhouse gas (GHG) emissions towards zero within their life cycle. This includes not only the fuels consumption in the neighborhood during the operation part but also the construction and deconstruction phase as well as the materials of the Neighborhood. The work of the center is pluri-disciplinary with, among others, work on architecture, energy system, materials and user behavior.
In this center, a software for minimizing the investment and operation costs of the energy system of ZENs is developed. It aims at helping stakeholder make decisions about the design of the energy system regarding sizing and choice of technologies in order to be a ZEN.

The main question behind this study is to assess whether and in what way the design of ZEN is affected by grid tariff design. This study is of particular interest for TSOs, DSOs and regulators because ZENs (or local systems based on similar concepts) are expected to be an important class of prosumers with high amounts of installed photovoltaic (PV) leading to potentially large imports of electricity in the winter and exports in the summer. This means they are a good subject for testing different grid tariffs structures and their impact on the neighborhood's import/export profiles of electricity.

\section{State of the Art and Contribution}

In the introduction, the traditional approach to grid tariffs and the way it is implemented in Norway was presented. This traditional approach is being challenged in the literature by some authors who think it is not suited for the current system or in the near future.
The reason that comes up the most often for justifying the need to change the tariffs is the emergence of prosumers and distributed generation in general. An increasing share of consumers are becoming producers of electricity and change the way the grid is used, which calls for a better allocation of costs or savings \cite{picciariello2015}, \cite{PICCIARIELLO201523},  \cite{SCHITTEKATTE2018}, \cite{FRIDGEN2018}.

Reference \cite{picciariello2015} discusses the need for new tariff design methodologies because of the growth of distributed generation. He identifies that the current challenges are the exemptions from tariffs for distributed generation and the volumetric tariffs with net metering; where in both cases the pricing does not represent the cost structures of DSO with high fixed cost and low variable costs. He also reviews different proposals of new tariffs structure.
Reference \cite{PICCIARIELLO201523} tackles the problematic of cross-subsidization between consumers and prosumers when net metering is used and suggest a cost-causality tariff structure. He highlights that the cross subsidization problem is particularly pronounced with net metering and energy based tariffs.
Reference \cite{FRIDGEN2018} studies the impact of different grid tariffs on residential microgrids. The grid tariffs were a combination of different volumetric tariff share on top of flat, time of use, critical peak or real time structures. He found that volumetric tariffs are more expensive for the consumers and lead to sharp load and generation peaks while the opposite is true when the tariff is not energy based.

Reference \cite{SCHITTEKATTE2018} analyzes the effects of different grid tariffs against different scenarios for the price of batteries and of PV. He warns against the possibility of distorting investment decision in case of poorly chosen grid tariff.

Reference \cite{honkapuro2017} study the opportunity for a new grid tariff structure in Finland for small scale customers, in particular incorporating a power part, and find it performs better with regard to cost-reflectivity and incentivizes consumers to be flexible.

Dynamic tariffs is one of the tariff structures that could be a possible improvement over the current grid structure. However other problems could arise such as fairness or cost recovery. Reference \cite{NEUTELEERS2017} studied the fairness of dynamic grid tariffs and pointed that it is important to remember all the principles of tariff design when assessing them.

Several studies looked into the relation between grid tariffs and prosumers with PV and batteries. Reference \cite{Bremdal18} uses measured data and simulation to show that in the Nordic countries, a power component in the tariff would be beneficial but the PV would still not allow to reduce the peak load. 
Reference \cite{JARGSTORF201585} takes into consideration the user reaction to the tariff with regard to self-consumption when assessing several grid tariffs based on capacity pricing.
Similarly, \cite{schreiber13} proposes a capacity based tariff, increasing quadratically with respect to power and linearly with energy and updated every 15 minutes, to allow the PV and battery system to benefit the grid in addition to the self-consumption. The optimized operation of the storage in addition to the capacity tariff allows a considerable reduction of peak imports and exports in exchange for only a small reduction of self consumption.

Few articles looked into different grid tariff structures applied to a model for investment in neighborhoods energy system. 
However, some studies have highlighted the impact in terms of investment in general of choosing a grid tariff structure. 
Reference \cite{firestone2006} showed that it is the fixed part of the cost that controls the amount of installed distributed generation and that the volumetric part has little influence on it. He suggests that public agencies can design countermeasures based on this result to obtain the desired amount of distributed generation. He also shows some results in terms of the chosen investments.
On the contrary \cite{abada_2018} warns against the risk of over investment by using a cooperative game theory approach to energy communities formation and investment in PV+battery system under different grid tariffs. They explain the over-investment observed in their results as a snowball effect due to the evolution of the grid tariffs as communities emerge and grid cost has to be recovered.
Reference \cite{SCHWARZ2018150} implements a purely capacity scheme and a specific dynamic scheme and compare the differences in terms of investment and several operation criteria such as annual self-consumption rate in a residential quarter.

To the authors knowledge, very few article have looked into the investment in neighborhoods energy system in order to look both at the change in investment and at the reaction of the neighborhood to the grid tariff in terms of operation, and none in a context of reduced green houses gases emissions such as ZEN.
This paper proposes to look into the impact of grid tariff design from two points of view. From a neighborhood planner perspective, how different grid tariff designs impact the investment choices. From a grid operator, how different tariff designs can shape the import and export of such neighborhoods and the revenue.
The neighborhood considered is zero emission in Norway and represent customers with a high amount of on-site production of electricity.

\section{ZENIT Model Description and Implementation}

ZENIT (Zero Emission Neighborhood Investment Tool) is a tool for minimizing the cost of investing and operating the energy system of a Zero Emission Neighborhood (ZEN). It uses a MILP model to find the optimal type and size of technology needed to provide heat and electricity to a ZEN.
The concept contains much more than only the energy system (materials and architecture to name two examples) but this tool's focus is energy systems.
The idea behind ZEN is to limit emissions and that it is possible to compensate for the various emissions of $CO_2$ in the neighborhood by exporting electricity to the grid. Indeed, by exporting electricity produced from on-site renewable to the grid, we assume that the production in the system is reduced by the corresponding amount and thus reduces the emission of the total system.

The model used in this paper is based on the model presented in \cite{pinel18}. In this section, the main elements of the model will be repeated and the differences with the model from \cite{pinel18} presented. For the details on the model not repeated, one can refer to that paper. Then the implementation and case chosen will be presented.
The optimization is written in Python and uses Gurobi as a solver. In this paper, we interpret the definition of a ZEN as a neighborhood that has 0 emissions over its lifetime, which is set to 60 years. However due to practical reasons and to reduce the computational time, different periods of the lifetime can be defined using one representative year for each. In this study we use a single period.
The different decision variables are the amount of investment in each technology for heat, power and energy storage as well as the operation related variables defined for each hour (e.g. amount of electricity produced, amount of fuel consumed).
Multiple constraints are used, to enforce the $CO_2$ limitations necessary in the ZEN context and to represent the operation of the neighborhood and in particular of each technology. It is important to note that part load limitations and start up/shutdown constraints are not implemented.
The objective function of the optimization is the following:

\textit{Minimize}:
\begin{multline}
    \sum_{i} C_i^{disc}\cdot x_i + b_{hg} \cdot C_{hg} + \frac{1}{\varepsilon^{tot}_{r,D}} \sum_{i} C_i^{maint} \cdot x_i\\+\frac{1}{\varepsilon^{tot}_{r,D}}\bigg(\sum_{t}\Big(\sum_{f}f_{f,t}\cdot P_{f}^{fuel} + (P_{t}^{spot}+P^{grid}\\+P^{ret})\cdot (y_{t}^{imp}+\sum_{est}y_{t,est}^{gb\_imp})-P_{t}^{spot}\cdot y_{t}^{exp}\Big)\bigg)
\end{multline}

Where $C_i^{disc}$ is the discounted investment cost in technology $i$ including re-investments and salvage value, $x_i$ the capacity of technology $i$, $b_{hg}$ the binary variable for investment in a heating grid, $C_{hg}$ the cost of the heating grid, $\varepsilon^{tot}_{r,D}$ the discount factor for the whole lifetime of the neighborhood with discount rate r, $C_i^{maint}$ the annual maintenance cost, $f_{f,t}$ the fuel consumption, $P_{f}^{fuel}$ the fuel price, $P_{t}^{spot}$ the electricity spot price, $P^{grid}$ and $P^{ret}$ the grid and retailer tariffs, $y_{t}^{imp}$ and $y_{t}^{exp}$ the import and export of electricity from the neighborhood and $y_{t,est}^{gb\_imp}$ the import of electricity to the storage. The subscript $t$ represent timesteps, $i$ the technologies with in particular $f$ for the technologies using other fuel than electricity and $est$ for energy storages.
It minimizes the cost of investing in the different technologies and the operating costs, fuels, electricity and O\&M costs and contains the costs of the heating grid and a binary associated with it that also gives access to technologies at a neighborhood level.

The most important constraint in the case of ZENs is the $CO_2$ balance \eqref{co2bal}, whose principle was explained earlier. In \eqref{co2bal}, $y_{t,est}^{gb\_imp}$, $y_{t,est}^{gb\_exp}$ and $y_{t,est}^{pb\_exp}$ are respectively the import and export on the grid side battery and the export from the on-site technologies producing electricity, $\varphi^{CO_2}_{e}$ and $\varphi^{CO_2}_{f}$ the $CO_2$ factors for electricity and other fuels, $\eta_{est}$ the efficiency of battery and $y_{t,g}^{exp}$ the export of electricity from the on-site technologies.

\begin{multline}
\label{co2bal}
    \sum_{t}((y_{t}^{imp}+\sum_{est}y_{t,est}^{gb\_imp}) \cdot \varphi^{CO_2}_{e}) \\+ \sum_{t}\sum_{f} (\varphi^{CO_2}_{f} \cdot f_{f,t}) \leq \sum_{t}(\sum_{est}(y_{t,est}^{gb\_exp}\\+y_{t,est}^{pb\_exp})\cdot \eta_{est} +\sum_{g}y_{t,g}^{exp}) \cdot \varphi^{CO_2}_{e}
\end{multline}

\begin{figure}[b]
    \centering
    \includegraphics[width=0.48\textwidth]{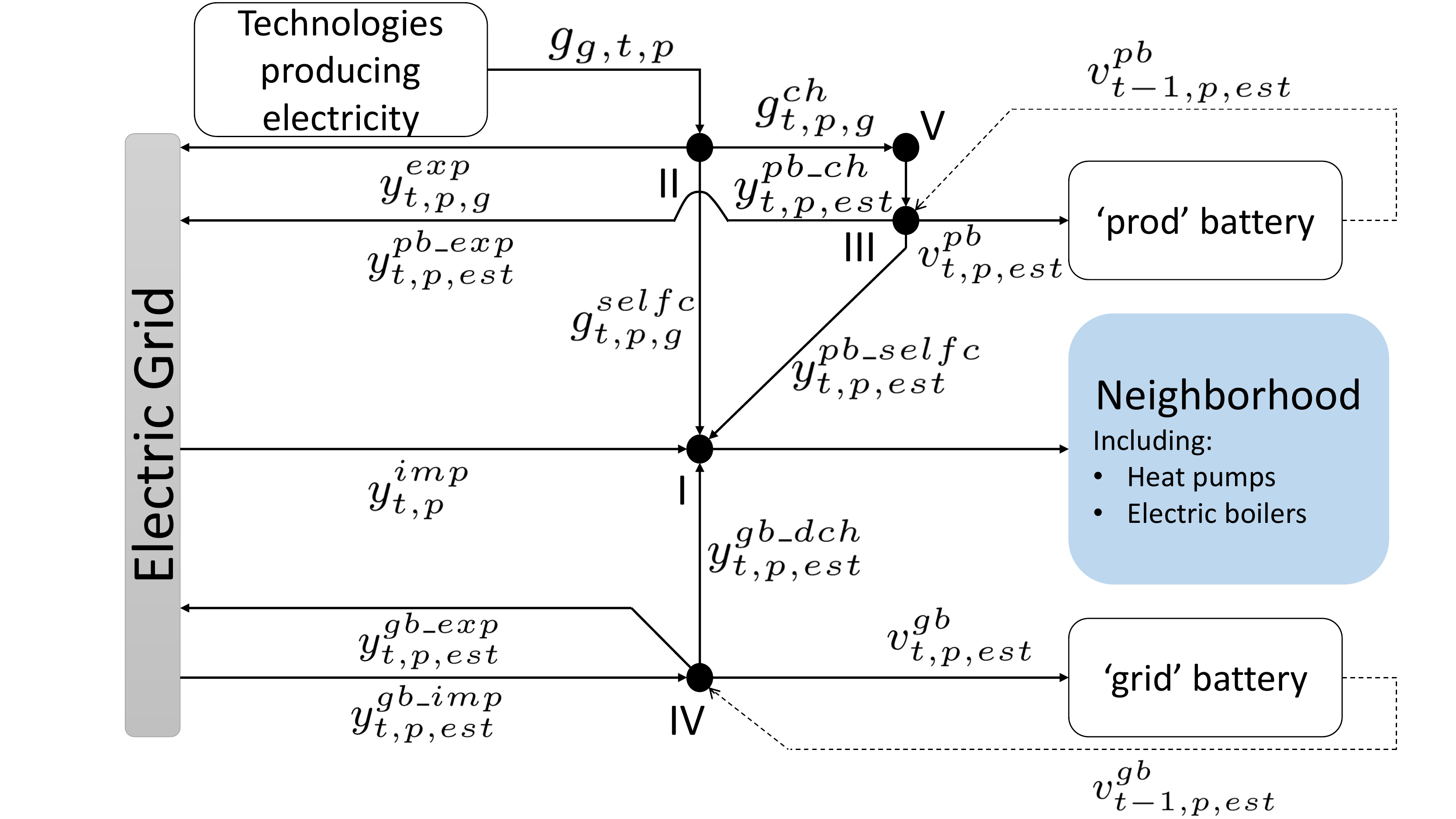}
    \caption{Graphical representation of the electricity balance in the optimization}
    \label{fig:elbal}
    \vspace{-1em}
\end{figure}

Fig. \ref{fig:elbal} presents graphically the electricity balance and the different equations associated.
Different technologies are included in the study; some of them are only available at the building level and others at the neighborhood level in a centralized production plant. The different technologies are: at the building level : Solar Panels (PV), Solar Thermal (ST), Heat Pumps (HP), Biomass Boiler (BB), Electric Boiler (EB), Gas Boiler (GB); and at the neighborhood levels: CHP (nCHP), Gas Boiler (nGB), Electric Boiler (nEB), Heat Pumps (nHP). In addition, Batteries (Bat) and Heat Storage (HS) are available at both levels. Different subcategories can be available to choose from within each category, for instance air-water or water-water heat pumps. In parenthesis is the notation used for the rest of the study for each technology.

Unlike the model in \cite{pinel18}, the model used for this paper uses a disaggregrated heat load. The buildings' load are not summed, but types of buildings are identified and the loads are aggregated per building type. It is possible to use a completely disaggregated heat load but the lack of available data motivated not doing it.

The input data necessary to run ZENIT are the electric and heating loads (ideally separated between domestic hot water and space heating), the outside and ground temperatures, the solar insolation and the electricity prices. Hourly timeseries for each representative years are necessary. A description of the neighborhood and its buildings with the floor and the roof area, and the layout of the neighborhoods is also needed.
In this study we assume the heating grid is there (and set the corresponding binary to 1) because there is one in the location that inspired this case. Its characteristics (layout, losses and cost) are then necessary but a module can be used to provide an estimate of the losses and cost based on the layout of the neighborhood. 
The $CO_2$ factors used were 17 g$CO_2$/kWh for electricity, 277 g$CO_2$/kWh for gas \cite{adapt13} and 7 g$CO_2$/kWh for wood chips \cite{dokka13}. The electricity produced via solar panels or solar thermal on-site do not have $CO_2$ associated. Embedded emissions were not included.
For additional details on the model and the references of the input data used in the model, refer to \cite{pinel18}.

\section{Grid Tariffs Description}

The Norwegian electricity consumption's recent trend is a consumption where peak demand increases relatively more than annual demand. This trend must be met by new incentives to shave peak load in order to avoid costly distribution grid investments. Grid tariffs are one effective way to solve this issue. In this paper we suggest three new grid tariffs and compare the results with the current grid tariff.

The first analyzed grid tariff is energy based and is the current tariff in Norway. It consists of an annual fixed price and a grid energy cost per kWh consumed. As this rate is flat, it does not incentivize flexible resources nor consumption patterns which results in lower peak demand. The annual cost can be calculated using \eqref{energytarifftotal}.

\begin{equation}
    C^{tot} = 137 + 0,0225\cdot \sum_t y_{t}^{imp\_tot}
    \label{energytarifftotal}
\end{equation}

The second grid tariff is a time-of-use based tariff which penalizes import when there is typically scarcity in the grid. The tariff has a basic cost, which is double during peak load hours (7-10am and 6-9pm) and reduced to half during low load hours (11pm-5am). The effect of increasing electric vehicle and demand response penetration on the peak hours is ignored. The total costs are given by \eqref{TOUtotal}.

\begin{multline}
    C^{tot} = \sum_t \big( 0,0123\cdot y_{t}^{imp\_low} + 0,0246\cdot y_{t}^{imp\_med} \\ 
    + 0,0492\cdot y_{t}^{imp\_peak} \big)
    \label{TOUtotal}
\end{multline}

The third tariff was originally described in \cite{Doorman}, and is called capacity subscription. It contains a fixed annual cost (\euro/year), a capacity cost (\euro/kW), an energy cost (\euro/kWh) and an excess demand charge (\euro/kWh). The energy cost is significantly higher when the imports are above the subscription. The main advantage of this tariff is that it incentivizes peak shaving and creates a market for capacity where consumers pay for the resource which in fact is scarce in the distribution grid: capacity. Disadvantages are complexity and the uncertainty in consumer behaviour. In addition, the optimal subscribed capacity is unknown in advance. Finding its value is further discussed in \cite{bjarghov}. In this paper, the subscribed capacity is a variable in the optimization. In reality, the consumer would have to choose it and it would most likely not be the optimal value.
%
The costs are calculated with \eqref{subcaptotal}.
\begin{equation}
   C^{tot} = 108 \cdot c^{sub} + \sum_t \big( 0,005\cdot y_{t}^{imp\_below} + 0,1\cdot y_{t}^{imp\_above} \big)
   \label{subcaptotal}
\end{equation}

The fourth tariff is a dynamic tariff where grid scarcity is taken into account. As an extra incentive to reduce impacts on the grid, a penalization $C^{sc}$ is given for consumption in hours with grid scarcity. Scarcity $\delta _{t}^{sc}$ in the system is defined as the 5\% of hours in the region (NO1) when the load is the highest. The percentage chosen is arbitrary and could be tuned or changed into a threshold by the regulator. The total costs are given by \eqref{dynamictotal}. In addition, as an added incentive to help the grid, a bonus for exporting in those hours is added, at the same cost as the scarcity tariff. In \eqref{dynamictotal}, $\delta _{t}^{sc}$ is a binary parameter defining for each hour if there is scarcity in the grid.
\begin{multline}
    C^{tot} = \sum_t \Big( \big( 0,0225\cdot(1-\delta _{t}^{sc}) + \delta _{t}^{sc} \cdot 0,1 \big) \cdot y_{t}^{imp\_tot} \\ - 0,1 \cdot  \delta _{t}^{sc} \cdot y_{t}^{exp\_tot} \Big)
    \label{dynamictotal}
\end{multline}

\section{Results}

In Norway, the legislation regarding prosumers is changing, moving from a situation where exports are limited to 100kW to a situation of unrestrained export. For this reason, both cases are investigated to explore the consequences on the design of ZENs of the different grid tariffs in these cases.
The investment in the energy system can be seen in Table \ref{tab:inv_yearly} and in Table \ref{tab:inv_yearlyexplim}, respectively for the case without and with limitation on exports. The results are presented in the format Prod Plant/ Student Housing/ Normal Offices/ Passive Offices.

\begin{table}[b]
\renewcommand{\arraystretch}{1.15}
    \centering
    \caption{Change in Investment Between Energy Tariff Case and the Other Grid Tariffs. Format: (Production Plant/)Student Housing/Normal Offices/Passive Offices}
    \begin{tabular}{@{} l r r r r @{}}
    \hline
        \textbf{Tech.} & \textbf{Energy} & \textbf{ToU} & \textbf{Subscribed} & \textbf{Dynamic}\\
        \hline
        nPV (kW) & 298/298/298 & 299/298/299 & 298/298/298 & 298/298/299 \\
        HP (kW) & 148/0/0/14.7 & 144/0/0/14.7 & 151/0/0/14.3 & 150/0/0/14.2 \\
        nBB (kW) & 0/0/1.7 & 0/0/0.9 & 0/0/2.2 & 0/0/2.4 \\
        GB (kW) & 0/0/0/0.6 & 3,1/0/0/3.7 & 0/0/0/2.3 & 0/0/0/2.5 \\
        HS (kWh) & 27/119/... & 81/104/... & 25/114/... & 49/134/... \\
         & ...47/27 & ...69/28 & ...122/33 & ...71/31 \\
         \hline
    \end{tabular}
    \label{tab:inv_yearly}
\end{table}

\begin{table}[b]
\renewcommand{\arraystretch}{1.15}
    \centering
    \caption{Change in Investment Between Energy Tariff Case and the Other Grid Tariffs, with Export Limits}
    \begin{tabular}{@{} l r r r r r @{}}
    \hline
        \textbf{Tech.} & \textbf{Energy} & \textbf{ToU} & \textbf{Subscribed} & \textbf{Dynamic} \\
        \hline
        nPV (kW) & 411/411/411 & 412/412/412 & 410/410/410 & 411/411/410\\
        HP (kW) & 147/0/0/14.6 & 147/0/0/14.6 &147/0/0/14.7 & 147/0/0/14.6\\
        EB (kW) & 89.2/0/0/0  & 88.7/0/0/0 & 88.4/0/0/0 & 89.9/0/0/0\\
        HS (kWh) & 320/324/... & 320/324/... & 335/323/... & 323/324/...\\
         & ...227/74 & ...227/72 & ...227/72 & ...227/78\\
        Bat (kWh) & 1774 & 1539 & 1519 & 1540\\
         \hline
    \end{tabular}
    \label{tab:inv_yearlyexplim}
\end{table}

The investments stay similar, no new technology is introduced or replaced. However, small variations in the amount of each technology appear, in particular heat storage. The difference between the energy system with and without export limit is greater, namely due to storages. A large battery pack is necessary in order to store the PV production while it waits to be exported, i.e. to accommodate the bottleneck. In addition, large investments in heat storages and electric boilers are done.
The subscribed capacity resulting of the optimization is of 134,5kW for the case with no export limit, and of 124kW in the case with export limits.

\begin{figure}[b]
    \centering
    
    \subfloat[Total Cost of the Neighborhood Energy System]{
    \includegraphics[width=0.23\textwidth]{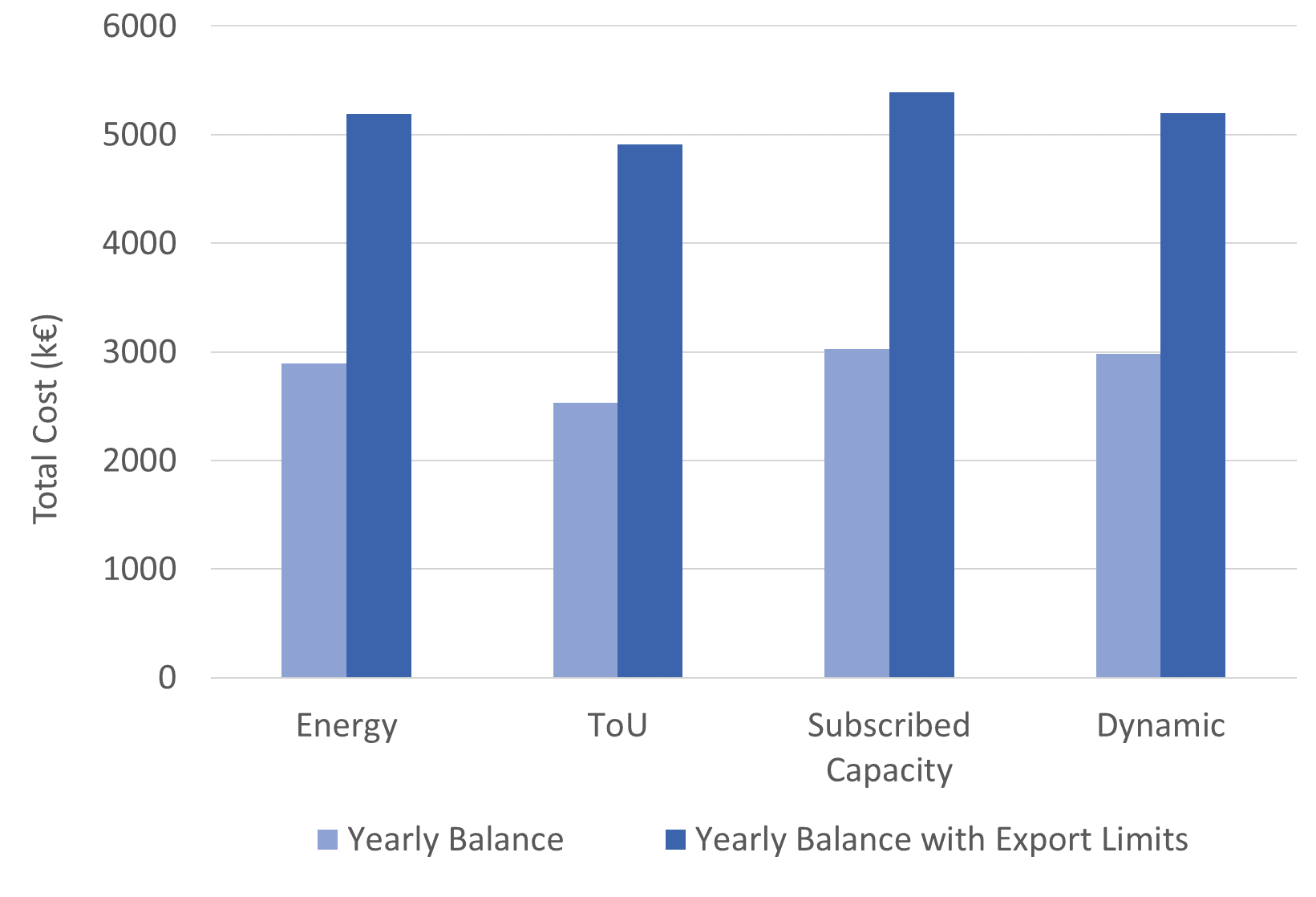}
    }
    \subfloat[Total Revenue of the DSO]{
    \includegraphics[width=0.23\textwidth]{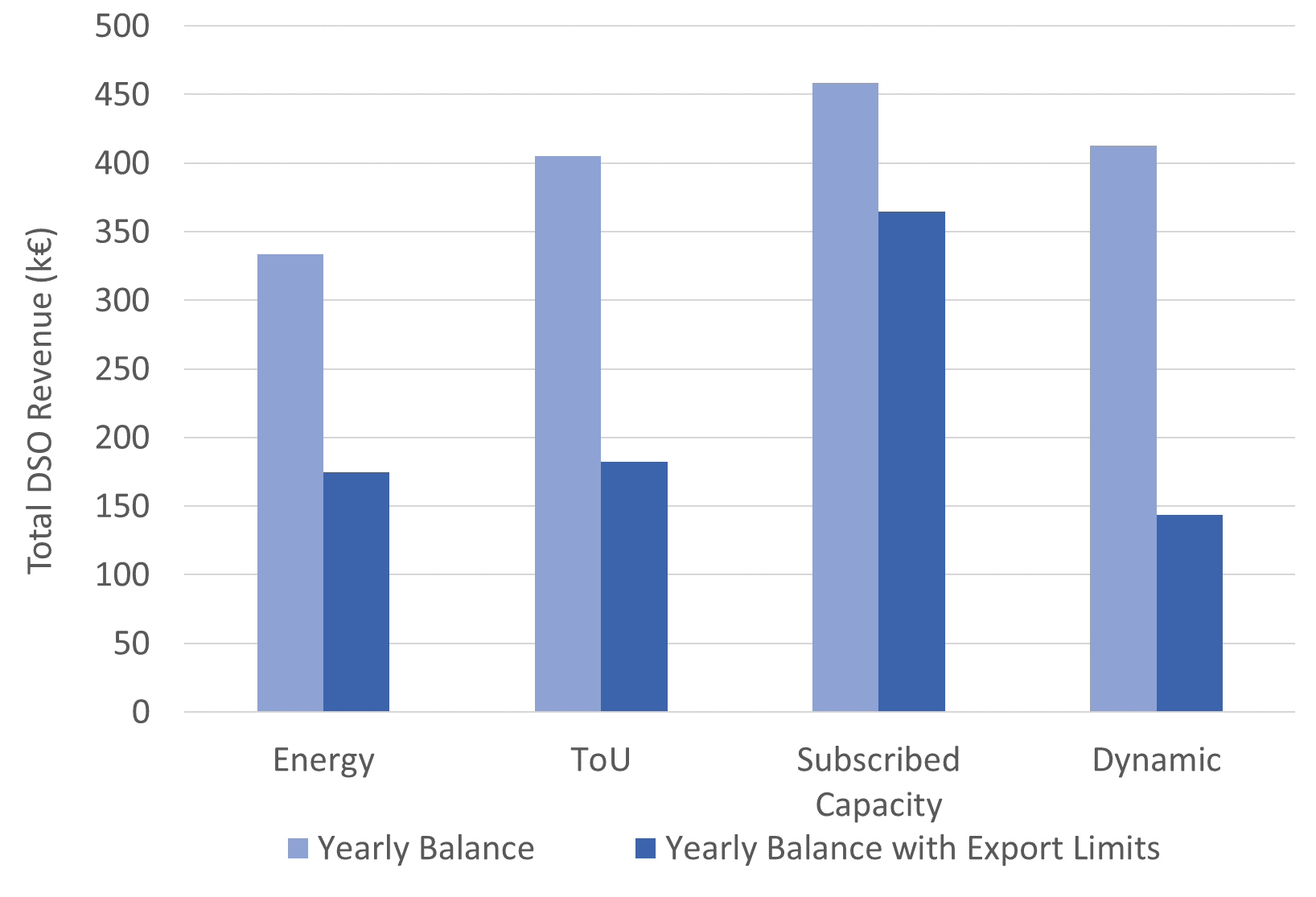}
    }
    \caption{Cost and DSO Revenue, Discounted to the Start}
    \label{fig:investment}
    \vspace{-1em}
\end{figure}

Fig. \ref{fig:investment} presents the total cost of the neighborhood's energy system (investment and operation) and the total revenue for the DSO, both over the lifetime and discounted to the start of the study. 
There are small variations in the cost in all cases. Subscribed capacity and dynamic pricing cause an increase in the total cost for the ZEN between 3 and 5\% compared with the energy case. On the other hand, the time of use scheme allows for a cost reduction of around 12\% in the case without export limit and 5\% with export limit.

The DSO revenue from the ZEN are higher when using the other pricing schemes than with the energy scheme when there is no export limit. When there is export limits, the DSO revenue stays the same because the battery allows to self-consume more and "anticipates" the higher price periods and buys electricity when the price is lower. The revenue in the case of export limits are about half of the revenue of the case of no export limit except in the case of subscribed capacity where the subscription tariff allows to maintain the revenue. 
The cost increase in the ZEN is of the same order of magnitude as the increase in revenue for the DSO except for ToU where the cost of the ZEN decreases while the revenue for the DSO increases. ToU has a beneficial effect from both points  of view in this aspect.

\begin{figure}[t]
    \centering
    \subfloat[No Export Limit]{
    \includegraphics[width=0.23\textwidth]{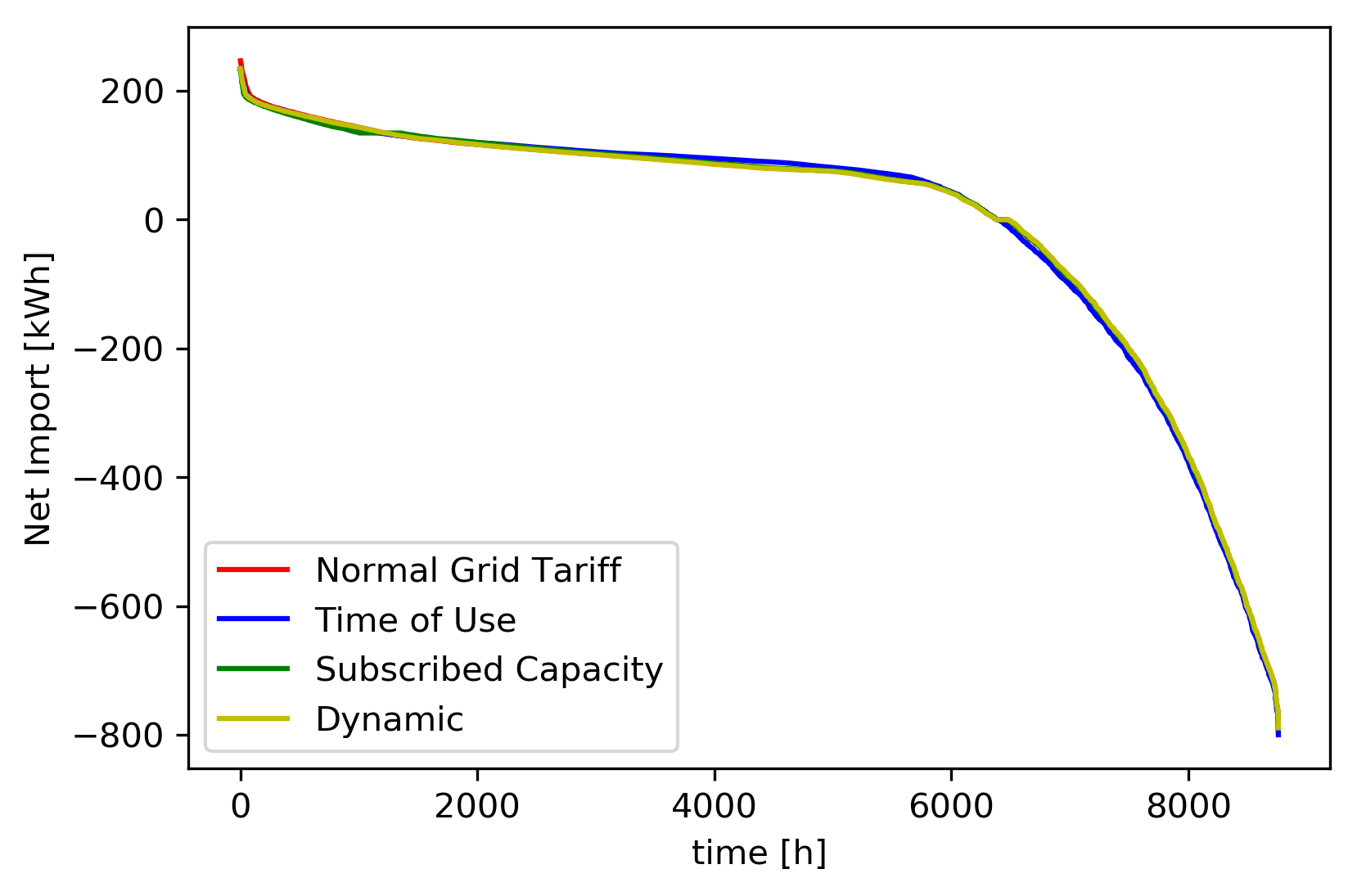}
    \label{subfig:noexp}
    }
    \subfloat[Export Limit]{
    \includegraphics[width=0.23\textwidth]{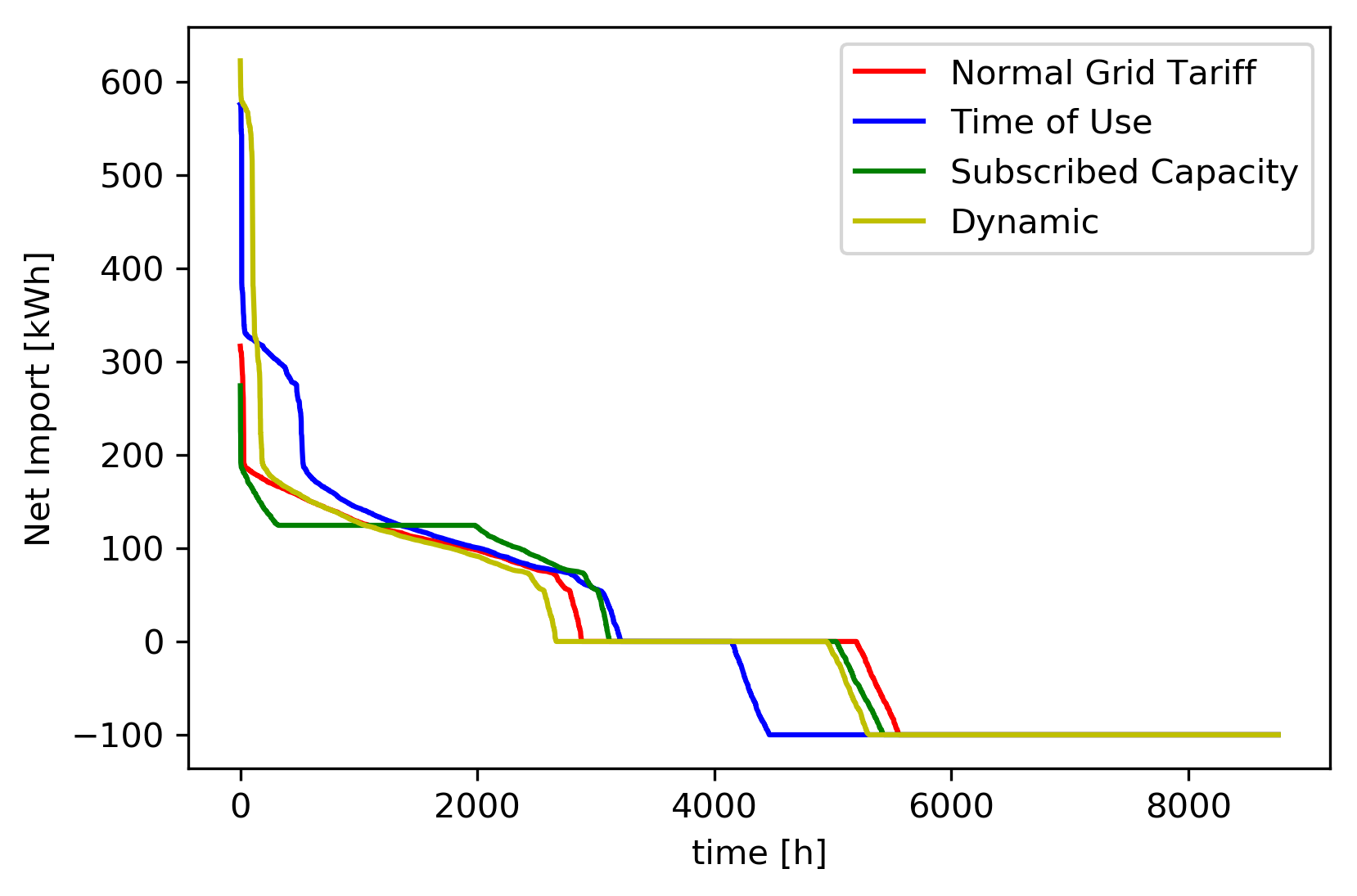}
    }
    \caption{Duration Curve of net Imports for the ZEN}
    \label{fig:duration}
    \vspace{-1.2em}
\end{figure}

The duration curves Fig. \ref{fig:duration}, in the case of no export limit, are not affected much by the tariff scheme in place. When export limitations are introduced, there are significant differences in the duration curves. The maximum imports from the ZEN are presented in Table \ref{tab:max_import}. ToU and dynamic schemes lead to really high imports, however they are not on peak hours but they still could cause congestion problems locally. In addition, ToU has a considerable number of hours with high loads of around 300kWh, which is not the case with the other schemes. On the contrary, subscribed capacity is able to keep imports below the subscribed capacity level most of the time.

\begin{table}[t]
\renewcommand{\arraystretch}{1.15}
    \centering
    \caption{Maximum Imports of Electricity}
    \begin{tabular}{@{} l r r r r @{}}
    \hline
        \textbf{Case} & \textbf{Normal} & \textbf{ToU} & \textbf{Sub. Cap.} & \textbf{Dynamic}\\
        \hline
        No Exp. Limits (kWh/h) & 246.6 & 234.9 & 231.3 & 234.6\\
        Exp. Limits (kWh/h) & 316.4 & 575.8 & 274.0 & 622.2\\
         \hline
    \end{tabular}
    \label{tab:max_import}
\end{table}

\begin{figure}[b]
    \centering
    \includegraphics[width=0.48\textwidth]{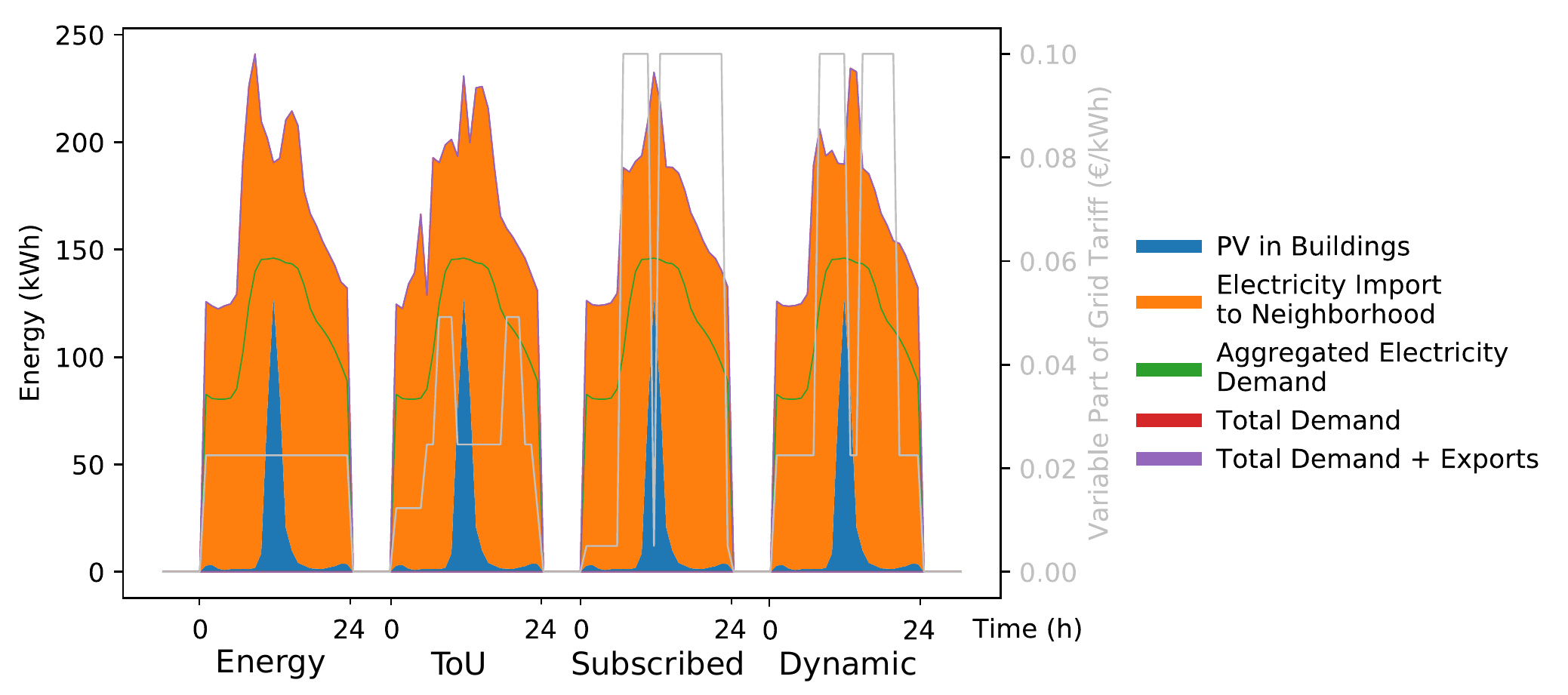}
    \caption{Operation of ZEN in a day in winter in the case of no export limit}
    \label{fig:dayopnoexplim}
    \vspace{-1em}
\end{figure}

In the case of no export limit, the operation is not affected much. However subscribed and dynamic allow to remove the peak import by shifting loads. On Fig. \ref{fig:dayopnoexplim}, for subscribed and dynamic, it seems that there is a peak in mid day but it is simultaneous with a peak in PV production, so the overall import profile is quite flat. However for the other pricing schemes the peak of PV production is decoupled from the peak in imports, which means that the peak remains, with a large dip in between them. This effect probably mitigates depending on the time of the year, since the duration curves on Fig. \ref{subfig:noexp} are almost the same.

\begin{figure}[b]
    \centering
    \includegraphics[width=0.48\textwidth]{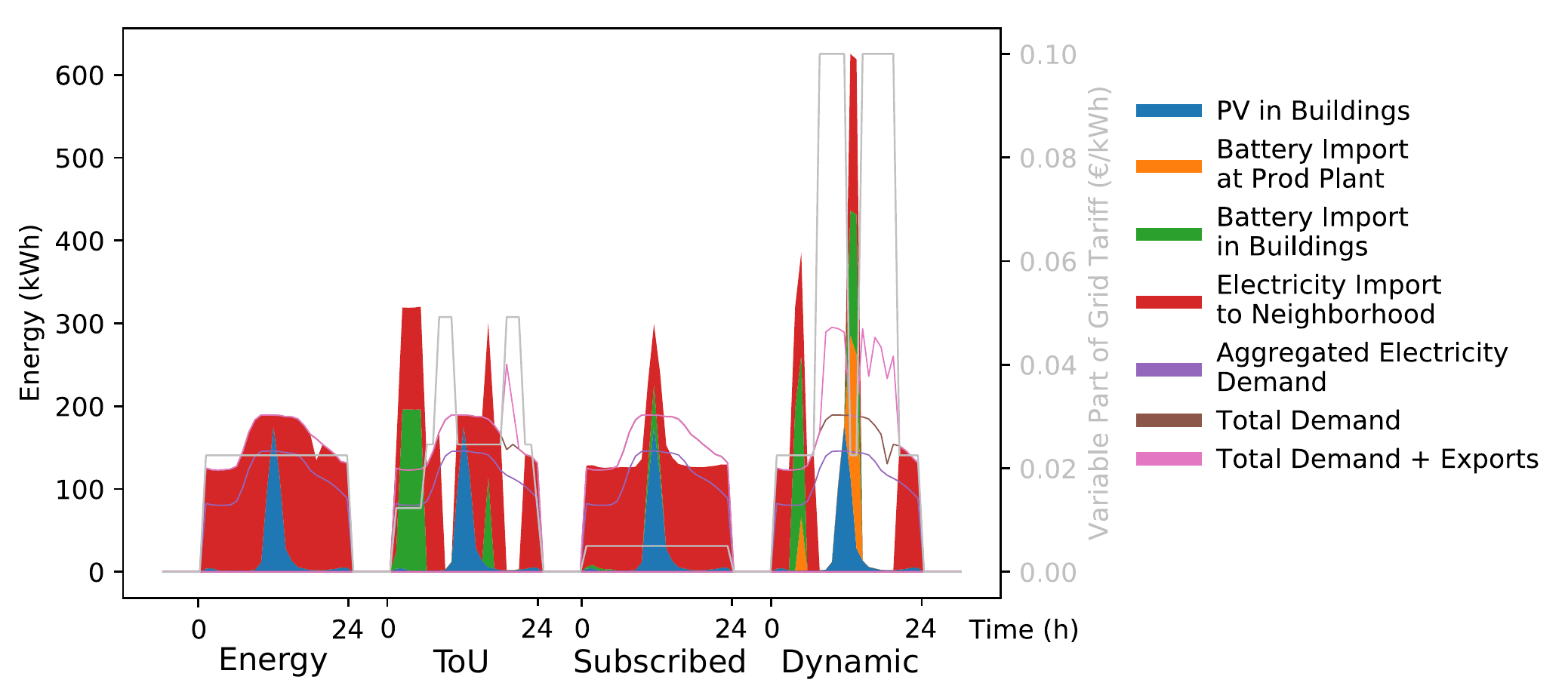}
    \caption{Operation of ZEN in a day in winter in the case of export limit}
    \label{fig:dayoplyexplim}
    \vspace{-1em}
\end{figure}

In the case of export limit Fig. \ref{fig:dayoplyexplim}, the batteries that are part of the system allow for more variations depending on the tariff scheme. In the energy scheme, the battery is used very little. In the ToU scheme, the optimization takes advantage of the low price hours to store energy in the battery and use it in the high price hours. It results in a higher load early in the morning which is most likely not problematic for the grid. In the subscribed capacity scheme, the battery is used to limit to the minimum the import above the subscribed capacity limit. During the peak of PV production, the battery imports from the grid because it is now below the subscription limit again. 

In the dynamic scheme, some hours of the day have an activation, meaning that they are part of the 5\% highest load in the year. The tariff in that case are extremely high and the battery is used as much as possible in those time periods, there is no import and the grid is relieved which was the intention behind using this scheme. However it also translates in high peaks when there is no activation, in order to fill the battery before the next one. This effect creates huge peak imports and one can wonder if the grid would be able to cope with them. There is no activation so they are not part of the 5\% highest load but there might still be an important load and this high peak creates congestion. Thus in the case of ZENs or highly flexible systems, such a dynamic pricing scheme could have unintended side effects.

\section{Conclusion}

Both from the DSO perspective and from the ZEN planner perspective, the results are quite dependant on the existence of export limits. Without export limits, it appears that the DSO could increase its revenue from new tariffs but those would translate as new cost to the ZEN. The exception is with the time of use tariff which is beneficial for both sides. The peaks are not reduced much by any new scheme and they are even higher in the case of export limitations.
In the case where export limits are set, the subscribed capacity scheme allows to preserve the revenue for the DSO, and offers reductions both in the peak and the number of hours with high imports. This tariff seems to be the most adapted to that case. From the ZEN perspective, this tariff is slightly more expensive but only because you do not profit from the reduction of the DSO revenue of the other tariffs. 
No matter the tariff implemented, the investments in the system with export limits are higher and costlier than when the export limit is not in place. 

The impact of grid tariff on ZEN is really dependent on the conditions for export. It can have very little effect or important impact both for the ZEN planner and for the DSO by simply modifying the conditions for export of electricity. Even though prosumers and consumers with high level of flexibility remain marginal in the grid, those effects should be taken into consideration while designing the tariffs and export conditions in order to maintain or offer a suitable environment for prosumers.

\section*{Acknowledgment}

This article has been written within the Research Centre on Zero Emission Neighbourhoods in Smart Cities (FME ZEN). The authors gratefully acknowledge the support from the ZEN partners and the Research Council of Norway.

\bibliographystyle{IEEEtran}
\bibliography{IEEEabrv,biblio}
\end{document}